\documentclass[preprint,12pt]{elsarticle}
\usepackage{graphicx} % Required for inserting images

\usepackage{amssymb}
\usepackage{amsmath}
\usepackage{multirow}
\usepackage[english]{babel}
\usepackage{array}
\usepackage{tabularx}
\usepackage{mathtools}
\usepackage{hyperref}

\begin{document}

\begin{frontmatter}

\title{ Plasma-Induced Modifications of the Shadows of Rotating Bardeen Black Holes with Perfect Fluid Dark Matter}

\vskip 1cm

\date{\today} %% Article title
%%             country={}}
\author[a]{\bf Gowtham Sidharth M\thanks{gowthamsidharth.m2019@vitstudent.ac.in}}
\author[b]{\bf Sanjit Das \thanks{sanjit.das@vit.ac.in}} %% Author name

%% Author affiliation

\affiliation{organization={School of Advanced Science ,Vellore institute of Technology},%Department and Organization
            addressline={}, 
            city={Chennai},
            postcode={}, 
            state={Tamilnadu},
            country={India}}

%% Abstract
\begin{abstract}
%% Text of abstract
We study the optical appearance of a rotating regular Bardeen black hole embedded in perfect fluid dark matter (PFDM) when photon propagation occurs through a plasma medium. Three plasma models are examined: a homogeneous distribution, a radially varying distribution, and a general distribution with both radial and angular dependence. The influence of plasma on photon motion and the resulting shadow morphology is analysed using shadow observables. To assess astrophysical viability, the plasma and PFDM parameters are constrained using the Event Horizon Telescope bounds on shadow circularity and fractional diameter deviation. The results demonstrate that environmental effects arising from both PFDM and plasma produce measurable modifications to the shadow, indicating that black-hole imaging can provide useful constraints on the surrounding medium as well as the intrinsic properties of regular rotating black holes.

\end{abstract}

%%Research highlights

%% Keywords

\end{frontmatter}
\section{Introduction}
\label{sec1}

Black holes that were once been a theoretical curiosity has now become an laboratories for testing gravity in the strong-field regime. Due to their intense gravitational curvature, these Black Holes influence the trajectory of photons in their vicinity. Hence the photon spends more time orbiting the black hole before either falling into the black hole or escaping to infinity.Thus creating a light ring boundary that separates the event horizon from rest of the outside spacetime. This shadow boundary is characterized by the internal geometry of the underlying spacetime.

The optical appearance of Schwarzchild black hole was studied by Synge \cite{Synge1950},.And later followed by Luminet to provide the earliest numerical simulations of an accreting black hole, illustrating how gravitational lensing shapes the observed image \cite{Luminet1979}. Following these early works on black hole shadows, more studies where done on rotating black holes, modified theories of gravity, and a variety of alternative compact-object models \cite{Bardeen1973,DeVries2000,Broderick2009,Grenzebach2015,Grenzebach2014,Chael2021,Gralla2020,Cunha2016,Moffat2015,Amir2018,Papnoi2014,Wei2013,Abdujabbarov2015}.
Further,The Event Horizon Telescope observations of the supermassive black holes M87* and Sgr A* have established black hole shadows as a direct observational window, opening an opportunity to explore theoretical models with astrophysical measurements.

General relativity predicts the existence of spacetime singularities within classical black hole solutions, where curvature invariant diverge and the classical description of gravity breaks down`. This limitation has motivated the search for singularity-free, or non singular regular, black hole solutions. The first such solution was proposed by Bardeen \cite{Bardeen1968}, and was later shown by Ayon-Beato and Garcia to arise from Einstein gravity coupled to nonlinear electrodynamics \cite{AyonBeato2000}. Due to its regular geometry and well-defined horizon structure, the Bardeen black hole has become one of the most extensively studied regular black hole models in gravitational physics.

The line element of static, spherically symmetric Bardeen spacetime is given by\cite{Bardeen1968}.
\begin{equation}
ds^{2}=-f(r)dt^{2}+\frac{dr^{2}}{f(r)}+r^{2}\left(d\theta^{2}+\sin^{2}\theta d\phi^{2}\right),
\end{equation}
where
\begin{equation}
f(r)=1-\frac{2Mr^{2}}{\left(r^{2}+g^{2}\right)^{3/2}},
\end{equation}
with $M$ denoting the black hole mass and $g$ representing the magnetic monopole charge associated with the nonlinear electrodynamics source.

A true cosmological study of astrophysical black holes should also account for the influence of their surrounding environment. Among the various astrophysical scenarios, dark matter is expected to play a significant role, particularly on galactic scales where supermassive black holes exists. Evidence for dark matter is provided by a wide range of astrophysical observations, such as galactic rotation curves, gravitational lensing phenomena, galaxy-cluster dynamics, and anisotropies in the cosmic microwave background. \cite{Rubin1980,Zwicky1937}. 

Various theoretical models have been developed to account for the nature of dark matter. These include the cold dark matter (CDM), warm dark matter, self-interacting dark matter, fuzzy dark matter, and scalar-field dark matter scenarios, each of which predicts different clustering behavior on small scales while remaining largely consistent with cosmological observations.\cite{Blumenthal1984,Moore1999,Bullock2017}..

Rather than focusing on the nature of dark matter,This study interested on the how the Dark matter affects the spacetime and photon propagation macroscopically. Hence, an alternative approach is to describe its gravitational influence through an effective energy--momentum tensor. Within this framework, Kiselev proposed the perfect fluid dark matter (PFDM) model, in which dark matter is represented as an anisotropic perfect fluid surrounding the black hole \cite{Kiselev2003}. The resulting spacetime admits an additional parameter that characterizes the dark matter distribution while preserving analytical tractability. The PFDM model has been widely employed to investigate black hole thermodynamics, gravitational lensing, quasinormal modes, and shadow formation \cite{Hou2018a,Haroon2019,Zhang2021}. This framework therefore provides a convenient setting for exploring how a surrounding dark matter environment modifies the optical appearance of black holes.

Another important factor that affects the photon propagation is the ionized plasma environment resulting from the black hole's  accreation disk.  Unlike in vacuum scenarios, the propagation of light in a plasma medium is affected by the refractive index that depends on both the plasma properties and the photon frequency. Hence in the presence of a plasma medium, this photon trajectory does not purely depend on null geodesics.The influence of plasma on gravitational lensing and black hole shadows has been investigated extensively in recent years. Perlick and Tsupko developed a general framework for describing light propagation in dispersive media within curved spacetimes \cite{Perlick2015,Tsupko2017}. These studies have demonstrated that the plasma distribution can produce measurable deviations from the corresponding vacuum shadow.

In this work, we examine the shadow cast by a rotating Bardeen black hole embedded in a perfect fluid dark matter (PFDM) halo and immersed in a non-magnetized plasma. Using the Hamilton--Jacobi formalism, we derive the equations governing photon trajectories and investigate how the magnetic monopole charge, spin, PFDM parameter, and plasma properties modify the shadow geometry. The resulting shadow observables are then confronted with Event Horizon Telescope (EHT) measurements to constrain the physically viable parameter space of the model.

The remainder of this paper is structured as follows. Section~2 introduces the spacetime geometry of the rotating Bardeen black hole in the presence of perfect fluid dark matter (PFDM). In Section~3, we incorporate the effects of a non-magnetized plasma and derive the corresponding photon equations of motion. Section~4 investigates the circular photon orbits, with particular emphasis on the co-rotating and counter-rotating trajectories in the equatorial plane. The influence of the spin parameter $a$, magnetic monopole charge $g$, PFDM parameter $\omega$, plasma strength $k$, and plasma frequency $\psi_{0}$ on the black hole shadow is analyzed in Section~5. The effective potential governing photon motion is discussed in Section~6. In Section~7, we compute the shadow radius $R_s$ and the corresponding angular diameter, and compare our results with the observational constraints from M87* to determine the allowed parameter space of the model. Finally, Section~8 summarizes the principal results and conclusions. Throughout this work, we adopt geometrized units with $G=c=1$ and normalize the black hole mass to $M=1$.
  
\section{Rotating Bardeen black hole in Perfect Fluid Dark matter}

The metric of a rotating Bardeen black hole in PFDM \cite{Zhang2021}  is obtained with the help of Newman - Janis algorithm\cite{Newman1965,Azreg2014},\\

\begin{equation}\label{e2}
\begin{split}
ds^2 & = -(1-\frac{2\rho r}{\Sigma})dt^2 + \frac{\Sigma}{\Delta_r}dr^2+ \Sigma d\theta^2 -\\& \frac{4 a \rho r \sin^2\theta}{\Sigma}dtd\phi -\sin^2\theta(r^2+a^2+\frac{2 a^2 \rho r \sin^2 \theta }{\sigma})d\phi^2
\end{split}
\end{equation}
with \\
\begin{equation}
2 \rho = \frac{2Mr^3}{(r^2+g^2)^{\frac{3}{2}}} -\omega \ln\frac{r}{|\omega|},
\end{equation}
\begin{equation}
\Sigma = r^2+a^2 cos^2\theta
\end{equation}
\begin{equation}
\Delta_r = r^2+ a^2 -\frac{2Mr^4}{(r^2+g^2)^{\frac{3}{2}}} + \omega r \ln\frac{r}{|\omega|}
\end{equation}

The logarithmic term with the parameter $\omega$ highlights how PFDM impacts the Bardeen metric. 

Following our previous study~\cite{Sidharth2026}, in which we established the two-branch horizon structure of the rotating Bardeen black hole in PFDM and analyzed the corresponding shadow properties and photon orbits, the present work extends the analysis by incorporating plasma effects. The PFDM parameter space, separated by the critical value $\omega_c$ due to the non-monotonic evolution of the outer event horizon, is adopted throughout the present investigation.

\section{Rotating Bardeen Black hole with Perfect Fluid Dark matter immersed in plasma}

Since most black holes are naturally surrounded by a high-energy, charged medium, it makes sense to consider our non-singular black hole, embedded in a PFDM environment, also immersed in a plasma. This surrounding plasma can, in general, be either magnetized \cite{Breuer1981} or non-magnetized \cite{Perlick2000}. For the sake of simplicity, we focus on a non-magnetized, pressureless plasma that does not interact with the PFDM. The propagation of photons in a plasma-filled curved spacetime is described by the Hamiltonian formulation originally proposed by Synge~\cite{Synge1960}. In the present analysis, we adopt the equivalent form developed by Perlick and Tsupko~\cite{Perlick2015}.
\begin{equation}\label{3.1}
H(x^{\mu},p_{\mu}) = \frac{1}{2}[g^{\mu \nu}p_{\mu}p_{\nu}+\psi_p(r)^2]
\end{equation}

The plasma frequency, denoted by $\psi_p(r)$, is assumed to depend only on the radial coordinate. In a dispersive plasma medium, the refractive index is determined by the local plasma frequency and the photon frequency $\psi$ measured by a static observer. It can therefore be written as \cite{Rogers2015}

\begin{equation}\label{3.2}
n^2(r,\psi)=1-\left(\frac{\psi_p(r)}{\psi}\right)^2.
\end{equation}

The photon frequency measured by the observer is related to its four-momentum through
\begin{equation}
\hbar\psi=-p_\mu u^\mu,
\end{equation}
where $u^\mu$ is the observer's four-velocity. For a static observer, the spatial components of the four-velocity vanish, yielding
\begin{equation}
\hbar\psi=-p_0u^0=-p_0\sqrt{-g^{00}}.
\end{equation}
Using this relation together with Eqs.~(\ref{3.1}) and (\ref{3.2}), the Hamiltonian describing photon motion in the plasma medium can be expressed as

\begin{equation}
H(x^{\mu},p_{\mu})=
\frac{1}{2}
\left[
g^{\mu\nu}p_{\mu}p_{\nu}
-
(n^2-1)
\left(-p_0\sqrt{-g^{00}}\right)^2
\right].
\end{equation}

Since the plasma frequency depends on the r, the plasma can be of many distributions, here we assume the distribution to be of radial power law, hence $\psi_p$ is expressed as\cite{Rogers2015}
\begin{equation}
\psi_p^2 = \frac{4\pi e^2 N(r)}{m_e}
\end{equation}

where $m_e$ is the mass of the electron with e being electronic charge. The plasma number density is assumed to follow a radial power-law profile~\cite{Synge1960,Rogers2015},
\begin{equation}
N(r)=\frac{N_0}{r^h},
\end{equation}
where $N_0$ sets the characteristic plasma density and $h\geq0$ determines the radial dependence of the distribution. Under this prescription, the refractive index is given by
\begin{equation}
n(r)=\sqrt{1-\frac{k}{r^h}},
\end{equation}
where $k$ characterizes the strength of the plasma medium. The parameter $h$ specifies the plasma profile: $h=0$ corresponds to a homogeneous plasma with a constant refractive index, whereas $h=1$ describes an inhomogeneous plasma whose refractive index decreases with radial distance. In the following analysis, both plasma configurations are considered.

In addition to these radial plasma profiles, we also examine a more general plasma configuration in which the plasma frequency depends on both the radial coordinate $r$ and the polar angle $\theta$.

\section{Black hole shadow}

The strong gravitational field surrounding a black hole significantly bends the trajectories of nearby photons. Depending on their initial conditions, some photons escape to distant observers, while others are captured by the event horizon. The unstable circular photon orbits define the critical curve that separates photons escaping to infinity from those captured by the event horizon. When projected onto the observer's sky, this critical curve forms the black hole shadow. For an observer at spatial infinity, the shadow boundary is described by the celestial coordinates $(\alpha,\beta)$,
\begin{equation}
\alpha=\lim_{r_0\rightarrow\infty}
-r_0^2\sin\theta_0
\left(\frac{d\phi}{dr}\right)_{(r_0,\theta_0)},
\end{equation}

\begin{equation}
\beta=\lim_{r_0\rightarrow\infty}
r_0^2
\left(\frac{d\theta}{dr}\right)_{(r_0,\theta_0)}.
\end{equation}
Here, $r_0$ denotes the distance between the observer and the black hole, while $\theta_0$ represents the inclination angle of the observer with respect to the black hole's rotation axis. The coordinate $\alpha$ gives the apparent displacement of the shadow along the horizontal direction, whereas $\beta$ gives the corresponding displacement along the vertical direction. By tracing photon trajectories reaching the observer and evaluating the corresponding celestial coordinates, the complete boundary of the black hole shadow can be constructed.
\subsection{Case I}\label{I}

We first consider a plasma medium whose refractive index varies only with the radial coordinate, i.e., $n=n(r)$. In this case, photon propagation through the rotating Bardeen spacetime is described by the Hamiltonian

\begin{equation}
\mathcal{H}
=\frac{1}{2}
\left[
g^{\mu\nu}p_\mu p_\nu
+(n^2-1)g^{00}p_0^2
\right].
\end{equation}

The corresponding photon trajectories are obtained using the Hamilton--Jacobi equation

\begin{equation}
\frac{\partial S}{\partial\lambda}
=
-\frac12
\left[
g^{\mu\nu}
\frac{\partial S}{\partial x^\mu}
\frac{\partial S}{\partial x^\nu}
+
(n^2-1)g^{00}
\left(
\frac{\partial S}{\partial x^0}
\right)^2
\right],
\end{equation}

together with the standard ansatz for the Jacobi action,

\begin{equation}
S=-Et+L_{\phi}\phi+S_r(r)+S_{\theta}(\theta),
\end{equation}

where $E$ and $L_{\phi}$ denote the conserved energy and angular momentum of the photon, respectively.

\subsubsection{Case Ia}

We first consider an inhomogeneous plasma with refractive index

\begin{equation}
n(r)=\sqrt{1-\frac{k}{r}}.
\end{equation}

Since the refractive index depends only on the radial coordinate, the Hamilton--Jacobi equation is not exactly separable. Following Refs.~\cite{11,c}, we therefore adopt the near-equatorial approximation by taking

\begin{equation}
\theta=\frac{\pi}{2}+\epsilon,
\end{equation}

where $\epsilon$ is a small perturbation. Throughout this work, the observer is assumed to lie in the equatorial plane, i.e., $\theta_0=\pi/2$.

Introducing the Carter constant $\kappa$, the radial motion is described by

\begin{equation}
R(r)
=
(n^2-1)E^2(r^2+a^2)^2
+\left[(r^2+a^2)E-aL_{\phi}\right]^2
-\Delta
\left[
(n^2-1)a^2E^2
+(aE-L_{\phi})^2
+\kappa
\right].
\end{equation}

The boundary of the shadow is determined by unstable circular photon orbits satisfying

\begin{equation}
R(r)=0,
\qquad
\frac{dR(r)}{dr}=0.
\end{equation}

Using these conditions, the impact parameter $\xi=L_{\phi}/E$ satisfies

\begin{equation}
A\xi^2+2B\xi+C=0,
\end{equation}

where

\begin{align}
A&=a^2\Delta', \nonumber\\
B&=2ar\Delta-a\Delta'(r^2+a^2), \nonumber\\
C&=
n^2\Delta'(r^2+a^2)^2
-
4r\Delta(r^2+a^2)\nonumber\\
&+
2nn'
\Delta^2a^2
-
(r^2+a^2)^2\Delta.
\end{align}

solving for $\xi$,the physically relevant solution is obtained by choosing the negative branch,

\begin{equation}
\xi
=
-\frac{B}{A}
-
\sqrt{
\left(\frac{B}{A}\right)^2
-
\frac{C}{A}
},
\end{equation}

The $\eta$ expression is calculated to be,

\begin{equation}
\eta
=
\frac{1}{\Delta}
\left[
(n^2-1)(r^2+a^2)^2
+
\left((r^2+a^2)-a\xi\right)^2
\right]
-
(n^2-1)a^2
-
(a-\xi)^2.
\end{equation}

Substituting these impact parameters into the expressions for the celestial coordinates yields

\begin{equation}
\alpha=-\frac{\xi}{n},
\qquad
\beta=\pm\frac{\sqrt{\eta}}{n}.
\end{equation}

The black hole shadow is obtained by plotting the celestial coordinates $(\alpha,\beta)$.

\subsubsection{Case Ib}\label{Ib}

We next consider a homogeneous plasma with a constant refractive index,

\begin{equation}
n=\sqrt{1-k}.
\end{equation}

Unlike the previous case, the Hamilton-Jacobi equation is completely separable, and the full equations of motion can be obtained without invoking the near-equatorial approximation. Proceeding in the same manner as in Case Ia, the radial and angular potentials are given by

\begin{equation}
R(r)
=
(n^2-1)E^2(r^2+a^2)^2
+\left[(r^2+a^2)E-aL_{\phi}\right]^2
-\Delta
\left[
(aE-L_{\phi})^2+\kappa
\right],
\end{equation}

\begin{equation}
\Theta(\theta)
=
\kappa
-(n^2-1)a^2E^2\sin^2\theta
+a^2E^2\cos^2\theta
-L_{\phi}^2\cot^2\theta.
\end{equation}

The boundary of the shadow is determined by the unstable circular photon orbits satisfying

\begin{equation}
R(r)=0,
\qquad
\frac{dR(r)}{dr}=0.
\end{equation}

The corresponding impact parameter $\xi=L_{\phi}/E$ satisfies

\begin{equation}
A\xi^2+2B\xi+C=0,
\end{equation}

where

\begin{align}
A&=a^2\Delta', \nonumber\\
B&=2ar\Delta-a\Delta'(r^2+a^2), \nonumber\\
C&=
n^2\Delta'(r^2+a^2)^2
-
4rn^2\Delta(r^2+a^2).
\end{align}

Solving the Quadratic equation, the expression for $\xi and \eta$ is obtained to be

\begin{equation}
\xi
=
-\frac{B}{A}
-
\sqrt{
\left(\frac{B}{A}\right)^2
-
\frac{C}{A}
},
\end{equation}

\begin{equation}
\eta
=
\frac{1}{\Delta}
\left[
(n^2-1)(r^2+a^2)^2
+
\left((r^2+a^2)-a\xi\right)^2
\right]
-
(a-\xi)^2.
\end{equation}

The celestial coordinates describing the black hole shadow are

\begin{equation}
\alpha
=
-\frac{\xi}{n}\csc\theta_0,
\qquad
\beta
=
\pm
\frac{
\sqrt{
\eta
-(n^2-1)a^2\sin^2\theta_0
+a^2\cos^2\theta_0
-\xi^2\cot^2\theta_0
}
}{n}.
\end{equation}

The shadow profile is obtained by plotting the celestial coordinates $(\alpha,\beta)$.
\subsection{Case II}\label{II}

We now consider the more general situation in which the plasma frequency depends on both the radial and angular coordinates, i.e., $n=n(r,\theta)$. In this case, the propagation of photons is governed by the Hamiltonian

\begin{equation}
\mathcal{H}(x^\mu,p_\mu)
=
\frac12
\left[
g^{\mu\nu}p_\mu p_\nu
+\psi_p^2(r,\theta)
\right],
\end{equation}

where the refractive index is given by

\begin{equation}
n^2(r,\theta)
=
1-
\left(
\frac{\psi_p(r,\theta)}{\psi}
\right)^2.
\end{equation}

The photon trajectories are obtained from the Hamilton--Jacobi equation using the standard ansatz

\begin{equation}
S
=
-Et
+
L_{\phi}\phi
+
S_r(r)
+
S_{\theta}(\theta).
\end{equation}

To achieve separability of the Hamilton-Jacobi equation, we adopt the plasma distribution proposed in Refs.~\cite{Perlick2017,Perlick2022},

\begin{equation}
\psi_p^2(r,\theta)
=
\frac{f_r(r)+f_{\theta}(\theta)}
{r^2+a^2\cos^2\theta},
\end{equation}

where $f_r(r)$ and $f_{\theta}(\theta)$ are arbitrary functions of the radial and angular coordinates, respectively.

Introducing the generalized Carter constant $\kappa$, the radial and angular potentials become

\begin{equation}
R(r)
=
\left[(r^2+a^2)E-aL_{\phi}\right]^2
-
\Delta
\left[
\kappa
+
f_r(r)
\right],
\end{equation}

\begin{equation}
\Theta(\theta)
=
\kappa
-
\left(
L_{\phi}\csc\theta
-
aE\sin\theta
\right)^2
-
f_{\theta}(\theta).
\end{equation}

The shadow boundary is determined by unstable circular photon orbits satisfying

\begin{equation}
R(r)=0,
\qquad
\frac{dR(r)}{dr}=0.
\end{equation}

Using these conditions, the impact parameter $\xi=L_{\phi}/E$ satisfies

\begin{equation}
A\xi^2+2B\xi+C=0,
\end{equation}

where

\begin{align}
A&=a^2\Delta', \nonumber\\
B&=2ar\Delta-a\Delta'(r^2+a^2), \nonumber\\
C&=\Delta'(r^2+a^2)^2
-4r\Delta(r^2+a^2)
+\Delta^2\tilde f_r'(r),
\end{align}

with

\[
\tilde f_r(r)=\frac{f_r(r)}{E^2}.
\]

The physically relevant solution corresponds to the negative branch,

\begin{equation}
\xi
=
-\frac{B}{A}
-
\sqrt{
\left(\frac{B}{A}\right)^2
-
\frac{C}{A}
},
\end{equation}

while

\begin{equation}
\eta
=
\frac{
\left[
(r^2+a^2)-a\xi
\right]^2
}{\Delta}
-
\tilde f_r(r).
\end{equation}

The celestial coordinates describing the shadow are therefore

\begin{equation}
\alpha
=
-\xi\csc\theta_0,
\qquad
\beta
=
\pm
\sqrt{
\eta
-
(\xi\csc\theta_0-a\sin\theta_0)^2
-
\tilde f_{\theta}(\theta_0)
},
\end{equation}

where

\[
\tilde f_{\theta}(\theta)
=
\frac{f_{\theta}(\theta)}{E^2}.
\]

To study the influence of different plasma distributions, we consider two representative models. The first assumes a purely radial plasma distribution,

\begin{equation}
f_r(r)
=
\psi_c^2\sqrt{M^3r},
\qquad
f_{\theta}(\theta)=0,
\end{equation}

while the second assumes a purely angular plasma distribution,

\begin{equation}
f_r(r)=0,
\qquad
f_{\theta}(\theta)
=
\psi_c^2M^2
(1+2\sin^2\theta).
\end{equation}

For each plasma model, the corresponding black hole shadow is obtained directly from the general expressions for the impact parameters and celestial coordinates derived above. The shadow profile is constructed by plotting the celestial coordinates $(\alpha,\beta)$ for an observer located at spatial infinity.

\section{Effect of Plasma and PFDM on Black Hole Shadows}\label{sec5}

We now investigate how a surrounding plasma modifies the shadow of a rotating Bardeen black hole in the presence of perfect fluid dark matter. The shadow characteristics are analyzed by varying the spin parameter $a$, magnetic monopole charge $g$, PFDM parameter $\omega$, plasma frequency, and plasma parameter $k$.

Figure~1 shows the black hole shadow for fixed values of the spin parameter, magnetic charge, and plasma parameter, while varying the PFDM parameter $\omega$. The analysis is carried out separately in the two allowed regions, namely $\omega<\omega_c$ and $\omega>\omega_c$. In the lower region ($\omega<\omega_c$), the shadow gradually shrinks as $\omega$ increases. In contrast, in the upper region ($\omega>\omega_c$), the shadow becomes larger with increasing $\omega$. This behavior is consistent with our previous study of the rotating Bardeen black hole in PFDM \cite{Sidharth2026}.

Figure~2 illustrates the shadow of the rotating Bardeen black hole for different values of the spin parameter while keeping the magnetic monopole charge fixed. Here, we consider the homogeneous plasma distribution (Case~I a) and vary the plasma strength parameter $k$. The shadow profiles are presented separately for the lower and upper PFDM branches. It is evident that the influence of the plasma on the shadow distortion is more pronounced in the lower PFDM branch than in the upper branch. Figure~3 presents the corresponding results for the inhomogeneous plasma distribution. Although the overall behavior remains qualitatively similar, the plasma-induced distortion is comparatively weaker. This is because the plasma density decreases with increasing radial distance in the inhomogeneous model, reducing the refractive influence on photon trajectories.

The black hole shadows of general plasma distribution where the plasma frequency depends on both r and $\theta$ are given in Figure 4 and 5. In both figure, the panels are arranged  for constant spin parameter a. The top row corresponds to lower magnetic monopole charge g,while bottom row for the for higher value of g. The left column represent lower bound PFDM range and right column shows the upper bound regime.

 Figure 4 shows the black hole shadow plots defined by plasma frequency with  \(f_r(r) = \psi_c^2 \sqrt{M^3 r}\) and \(f_{\theta}(\theta) = 0\). From the plots we inferred several key things, the distortion due to spin parameter a is more prominent in higher g values in lower PFDM range. In contrast, the spin-induced distortion is considerably weaker in the upper-bound PFDM regime.Furthermore, the plasma frequency has a weak effect on shape and size of the blackhole. 

Figure 5 shows the black hole shadow plots defined by plasma frequency with \(f_r(r) = 0\) and \(f_{\theta}(\theta) = \psi_c^2 M^2 (1 + 2 \sin^2 \theta)\),Unlike the previous plasma distribution, the angular plasma profile produces negligible changes in the shadow distortion but significantly affects its size. The size of the shadow decreases with increasing plasma frequency,this reduction in size is more pronounced in the upper bound PFDM range.

\begin{figure}
\label{f1.1}
\includegraphics[scale=0.7]{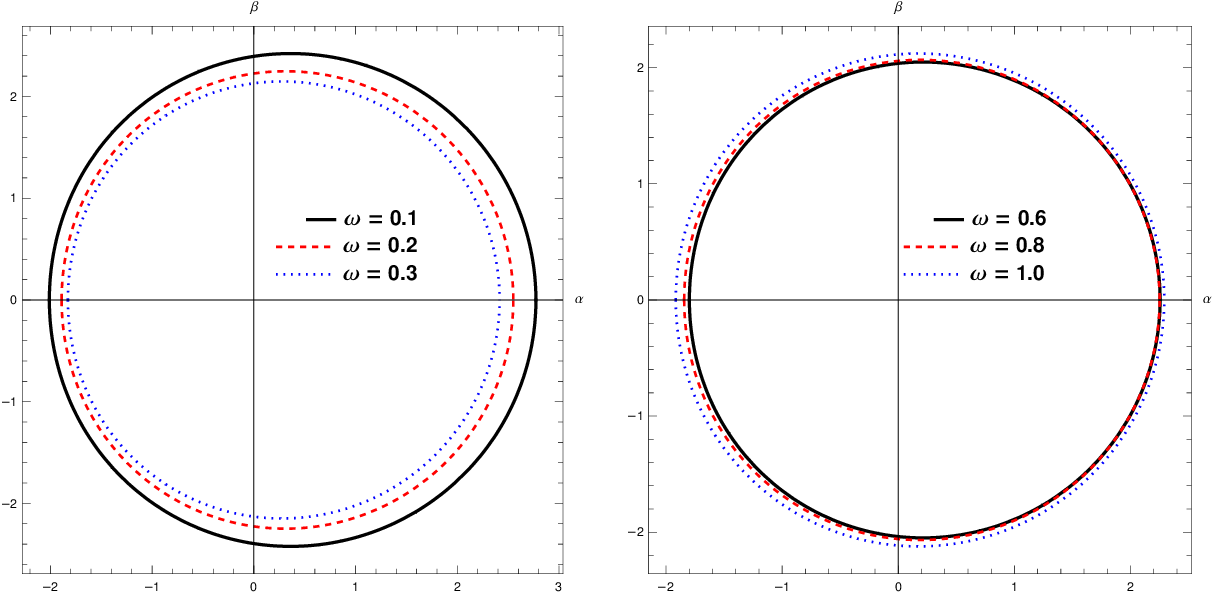}
\caption{Black hole shadow for different values of the PFDM parameter $\omega$, with $a=0.4$, $g=0.2$, and $k=0.2$. The left panel corresponds to the lower PFDM branch ($\omega<\omega_c$), while the right panel shows the upper PFDM branch ($\omega>\omega_c$).}
\end{figure}

\begin{figure}
\label{f1.1}
\includegraphics[scale=0.7]{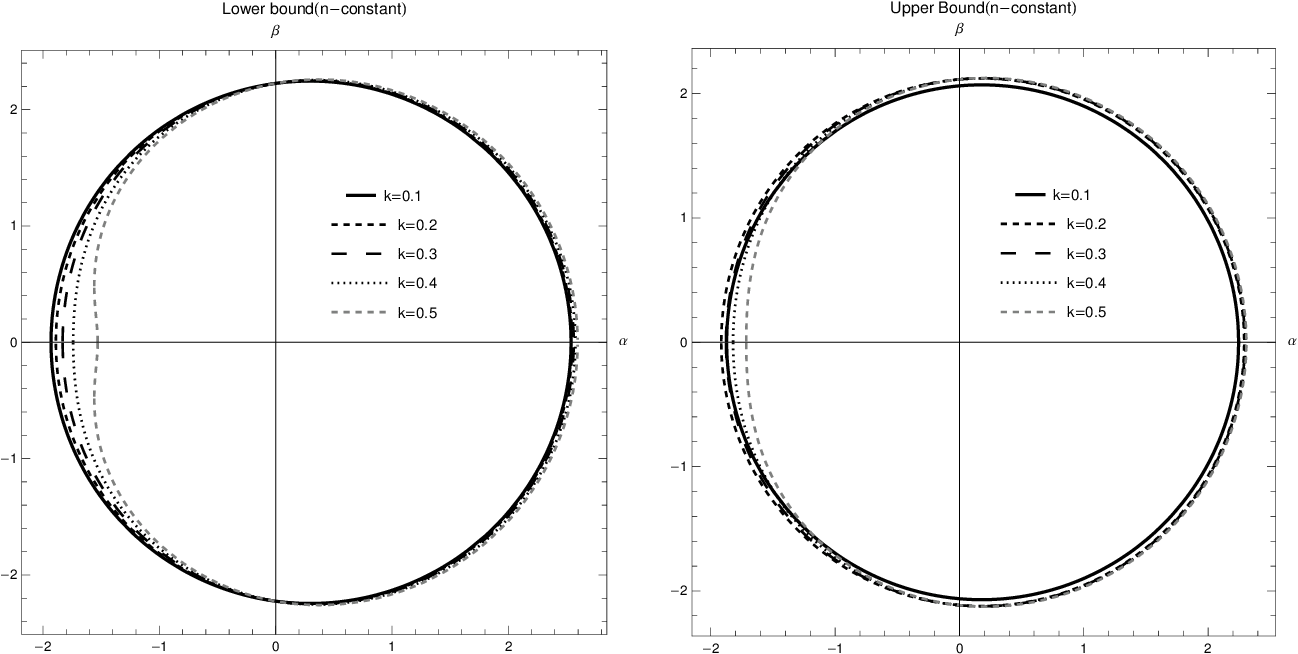}
\caption{Black hole shadow in a homogeneous plasma medium for different values of the plasma strength parameter $k$, with $a=0.4$ and $g=0.2$. The left and right panels correspond to the lower ($\omega=0.2$) and upper ($\omega=1$) PFDM branches, respectively.}
\end{figure}

\begin{figure}
\label{sh3}
\includegraphics[scale=0.8]{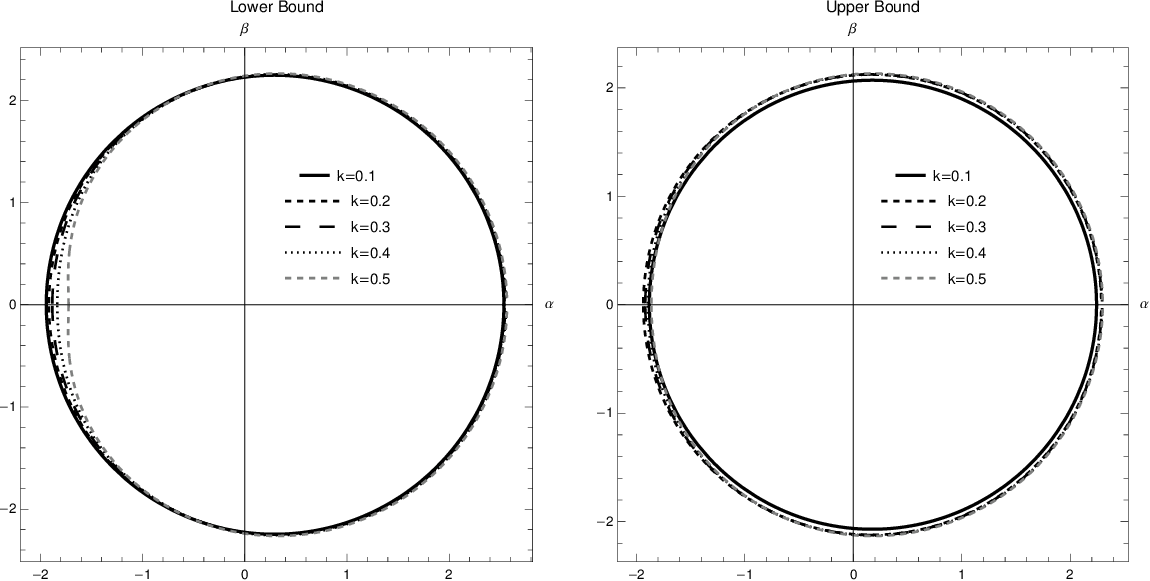}
\caption{Black hole shadow in a inhomogeneous plasma medium for different values of the plasma strength parameter $k$, with $a=0.4$ and $g=0.2$. The left and right panels correspond to the lower ($\omega=0.2$) and upper ($\omega=1$) PFDM branches, respectively.}
\end{figure}

\begin{figure}
\label{sh4}
\includegraphics[scale=0.9]{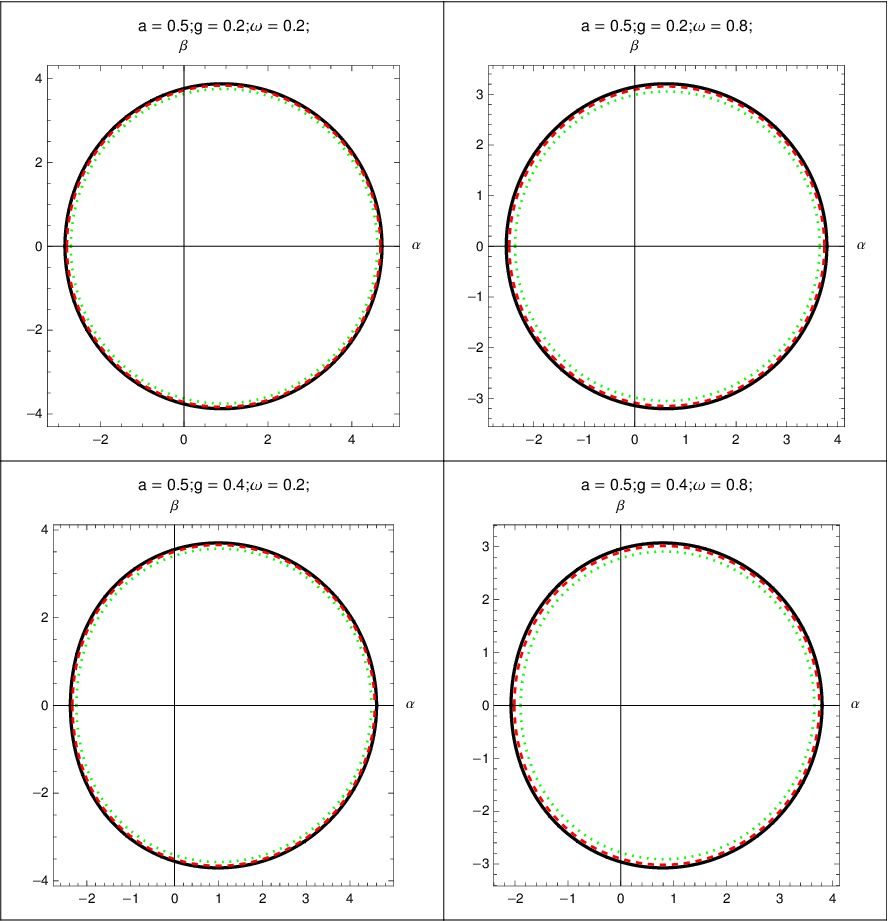}
\caption{Black hole shadow for different values of the normalized plasma frequency $\psi_c/\psi_o$. The top and bottom rows correspond to $g=0.2$ and $g=0.4$, respectively, while the left and right columns represent the lower ($\omega=0.2$) and upper ($\omega=0.8$) PFDM branches. The curves denote $\psi_c/\psi_o=0.0$ (black solid), $1.0$ (red dashed), $3.0$ (green dashed).}
\end{figure}

\begin{figure}
\label{sh5}
\includegraphics[scale=1.0]{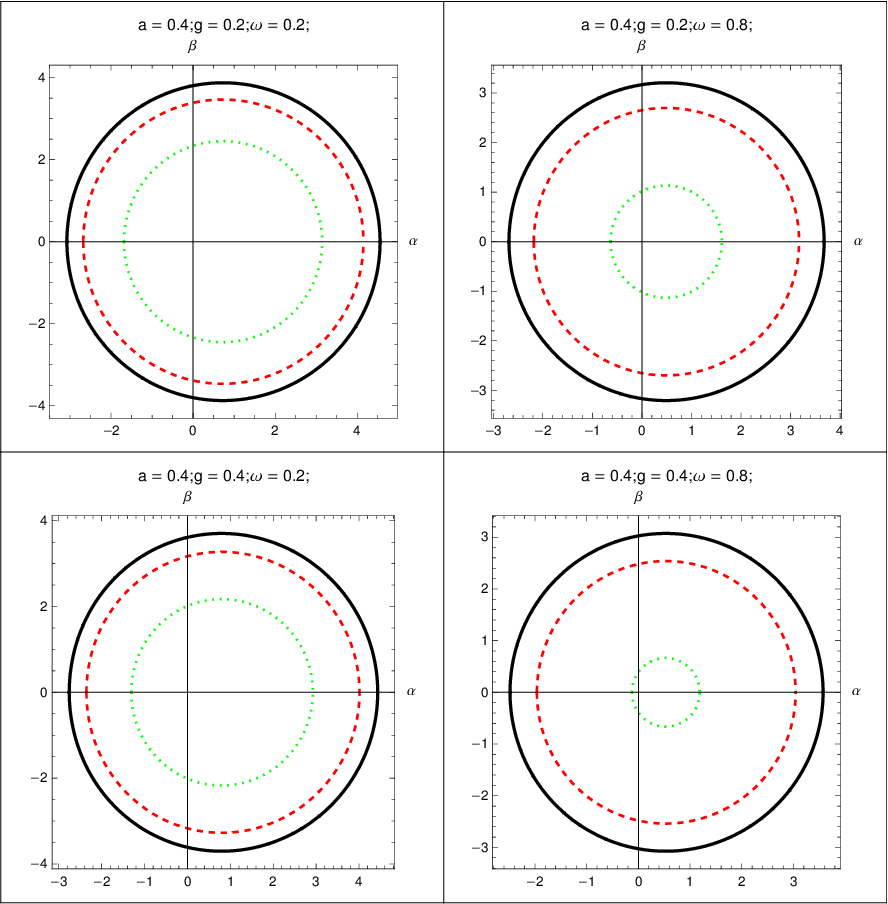}
\caption{Black hole shadow for different values of the normalized plasma frequency $\psi_c/\psi_o$. The top and bottom rows correspond to magnetic monopole charges $g=0.2$ and $g=0.4$, respectively. The left and right columns represent the lower ($\omega=0.2$) and upper ($\omega=0.8$) PFDM branches. The curves correspond to $\psi_c/\psi_o=0.0$ (black solid), $1.0$ (red dashed), and $3.0$ (green dashed).}
\end{figure}

\section{Effective Potential}

In this section, we analyze the effective potential, $V_{\mathrm{eff}}$, to study the nature of photon orbits around the black hole. The extrema of the effective potential determine the stability of the circular orbits. Stable circular orbits correspond to the minima of the effective potential, satisfying $\frac{\partial^2 V_{\mathrm{eff}}}{\partial r^2}>0$, whereas unstable circular orbits correspond to the maxima, satisfying $\frac{\partial^2 V_{\mathrm{eff}}}{\partial r^2}<0$. The effective potential is given by

\begin{equation}
\dot{r}^2 +V_{eff} = E
\end{equation}

where $\dot{r}$ is defined as

\begin{equation}
\dot{r} ^2= \frac{1}{r^4} \left[ (E^2(r^2+a^2)- a L)^2 -\Delta (a E -L)^2 +(n^2-1)(E^2(r^2+a^2)^2-\Delta E^2 a^2)\right]
\end{equation}

Figure 6 shows the variation of the effective potential in an homogeneous plasma distribution, $V_{\mathrm{eff}}$, with the radial coordinate $r$ for different values of the plasma parameter, while all other parameters are kept fixed. The analysis is carried out separately for the two allowed regions of the PFDM parameter, namely $\omega<\omega_c$ and $\omega>\omega_c$.

\begin{figure}
\label{fi6}
\includegraphics[scale = 0.5]{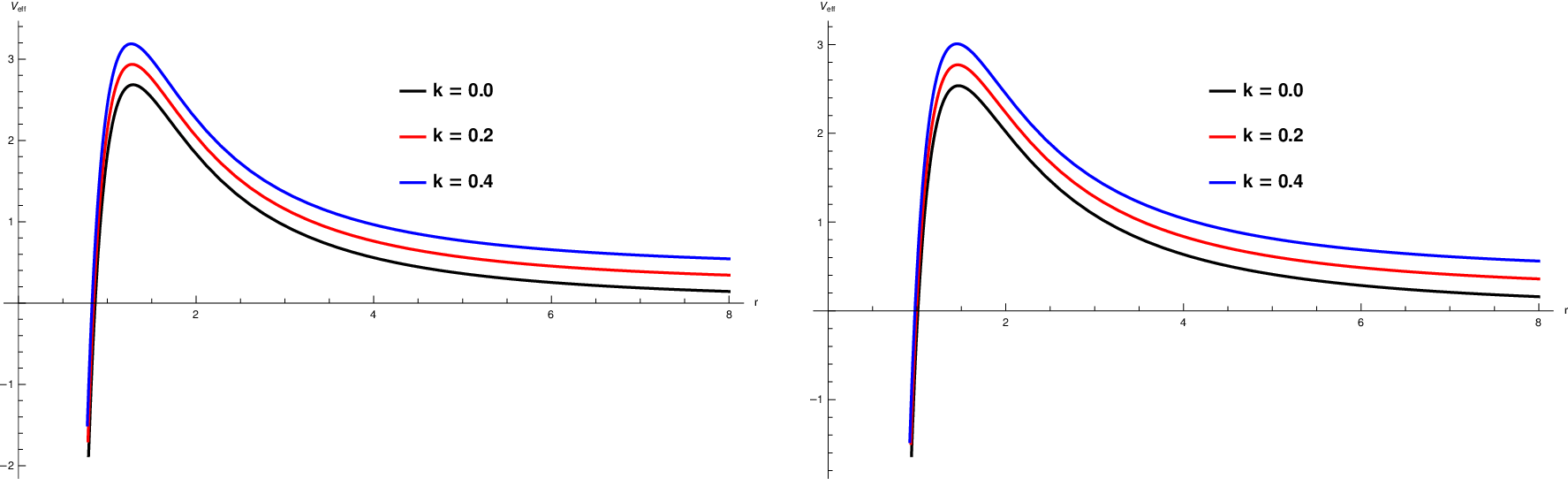}
\caption{Effective potential for photon motion in a homogeneous plasma medium for different values of the plasma strength parameter $k$, with $a=0.4$, $L=3$, and $g=0.5$. The left and right panels correspond to the lower ($\omega=0.2$) and upper ($\omega=1$) PFDM branches, respectively.}
\end{figure}

\begin{figure}
\label{fig7}
\includegraphics[scale = 0.6]{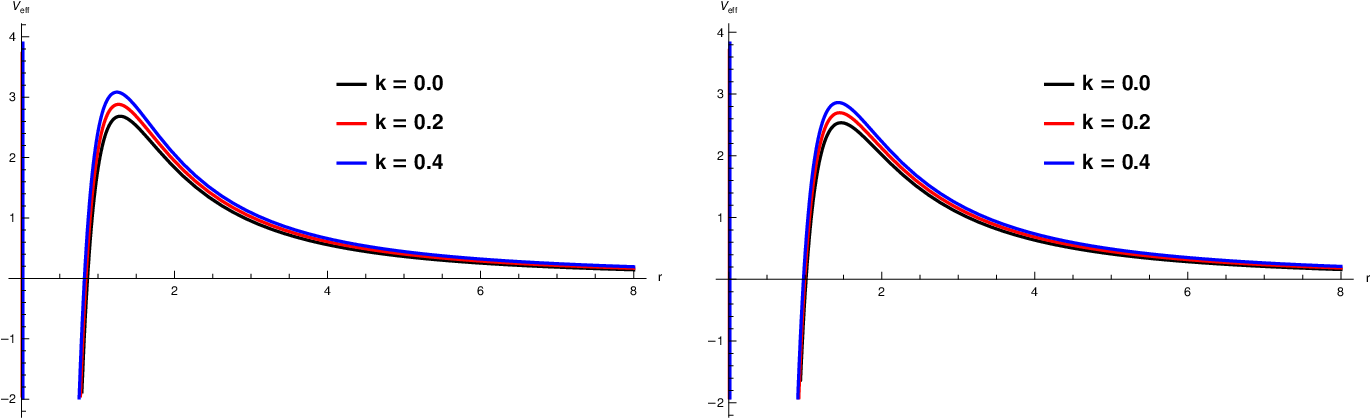}
\caption{Influence of the plasma strength parameter $k$ on the effective potential in an inhomogeneous plasma medium for $a=0.4$, $L=3$, and $g=0.5$. The left and right panels correspond to the lower ($\omega=0.2$) and upper ($\omega=1$) PFDM branches, respectively.}
\end{figure}

Figure 7 illustrates the effective potential experienced by photons orbiting the black hole for an inhomogeneous plasma distribution, where the refractive index is taken as

$n(r)=\sqrt{1-\frac{k}{r}}.$\\

The effective potential is plotted for different values of the plasma parameter $k$, while all other parameters are kept fixed. The left panel corresponds to $\omega=0.2$ ($\omega<\omega_c$), whereas the right panel corresponds to $\omega=1$ ($\omega>\omega_c$).

As the plasma parameter $k$ increases, the effective potential increases in both PFDM regions, exhibiting a trend similar to that observed for the homogeneous plasma distribution. The maxima of the effective potential correspond to unstable photon orbits. With increasing $k$, the peak of the effective potential shifts slightly towards smaller values of $r$, indicating a small inward shift in the radius of the unstable photon orbit.

The increase in the effective potential with the plasma parameter demonstrates that the plasma medium has a significant influence on photon propagation. As the plasma density increases, photons experience a stronger effective refractive medium, leading to modifications in their trajectories and, consequently, in the properties of the black hole shadow.

\section{Observational Constraints on Rotating Regular Black Holes in a Plasma Medium}

Observational measurements of M87* and Sgr A* provide an effective means of constraining the parameter space of rotating Bardeen black holes embedded in perfect fluid dark matter and surrounded by a plasma medium. To characterize the shadow geometry, we consider a shadow profile that is symmetric about the $\alpha$-axis, i.e., $\beta=0$. The geometric center of the shadow is determined from the area element $\mathcal{A}$ as

\begin{equation}
\alpha_c=\frac{\int \alpha\, d\mathcal{A}}{\int d\mathcal{A}}, \qquad
\beta_c=0.
\end{equation}

The radial distance from the shadow center to its boundary at an angular coordinate $\phi$ is defined by

\begin{equation}
\ell^2(\phi)=\left[\alpha(\phi)-\alpha_c\right]^2+\beta^2(\phi).
\end{equation}

The corresponding average shadow radius is given by

\begin{equation}
R_{\mathrm{avg}}^2=
\frac{1}{2\pi}
\int_{0}^{2\pi}
\ell^2(\phi)\,d\phi.
\end{equation}

Using the average radius, the deviation from circularity is evaluated as~\cite{Bambi,L1}

\begin{equation}
\Delta C=
\frac{1}{R_{\mathrm{avg}}}
\sqrt{
\frac{1}{2\pi}
\int_{0}^{2\pi}
\left[\ell(\phi)-R_{\mathrm{avg}}\right]^2
\,d\phi
}.
\end{equation}

Expressing $R_{avg}$ and $\Delta C$in $r_{ph}$ instead of $\phi$ for conviencce,
\begin{equation}
R_{\text{avg}}^2 = \frac{1}{\pi} \int_{r_{\text{ph}-}}^{r_{\text{ph}+}} \left(\beta'(\alpha - \alpha_c) - \beta \alpha'\right) dr_{\text{ph}},
\end{equation}
\begin{equation}
\Delta C = \frac{1}{R_{\text{avg}}} \sqrt{\frac{1}{\pi} \int_{r_{\text{ph}-}}^{r_{\text{ph}+}} \left(\beta'(\alpha - \alpha_c) - \beta \alpha'\right) \left(1 - \frac{R_{\text{avg}}}{\ell}\right)^2 dr_{\text{ph}}},
\end{equation}
and the geometric center,
\begin{equation}
\alpha_c = \frac{\int_{r_{\text{ph}-}}^{r_{\text{ph}+}} \alpha \beta \alpha' \, dr_{\text{ph}}}{\int_{r_{\text{ph}-}}^{r_{\text{ph}+}} \beta \alpha' \, dr_{\text{ph}}}, \beta =0.
\end{equation}

Here, $r_{ph^-}$ and $r_{ph^+}$ denote the radial coordinates at which the shadow boundary intersects the $\alpha$-axis. These points are obtained from the roots of the equation $\beta(r_{ph})=0$.

At present, the available EHT observations of Sgr A$^*$ do not provide meaningful constraints on the deviation from circularity, $\Delta C$. In contrast, observations of M87$^*$ place an upper bound of $\Delta C \lesssim 0.1$ for an observer inclination angle of $17^\circ$~\cite{L1,L5,L6,L12,L17}. This inclination is consistent with the orientation of the relativistic jet associated with M87$^*$.

Another quantity employed by the Event Horizon Telescope collaboration is the fractional deviation of the shadow diameter, denoted by $\delta$. It quantifies the relative difference between the shadow diameter predicted by a given black hole model and that of the Schwarzschild black hole.
\begin{equation}
\delta=\frac{d_{sh}}{d_{sh,sch}}-1=\frac{R_{sh}}{3\sqrt{3}M} -1
\end{equation}

Using two independent measurements of the mass and distance of Sgr A$^*$ from the VLTI and Keck observations, the EHT collaboration placed the following constraints on the fractional deviation parameter $\delta$ \cite{L12,L17}:

\begin{equation}
\delta = \begin{cases}-0.08^{+0.09}_{-0.09} & (VLTI)\\  -0.04^{+0.09}_{-0.10} & (Keck) \end{cases}
\end{equation}

Therefore, the VLTI and Keck observations constrain the fractional deviation parameter to the range $-0.14<\delta<0.01$. In addition, the observational data favor inclination angles below $50^\circ$, with values of $\theta_0>50^\circ$ being disfavored.

\begin{figure*}
     \centering
     \begin{minipage}{0.4\textwidth}
	\includegraphics[scale =0.6]{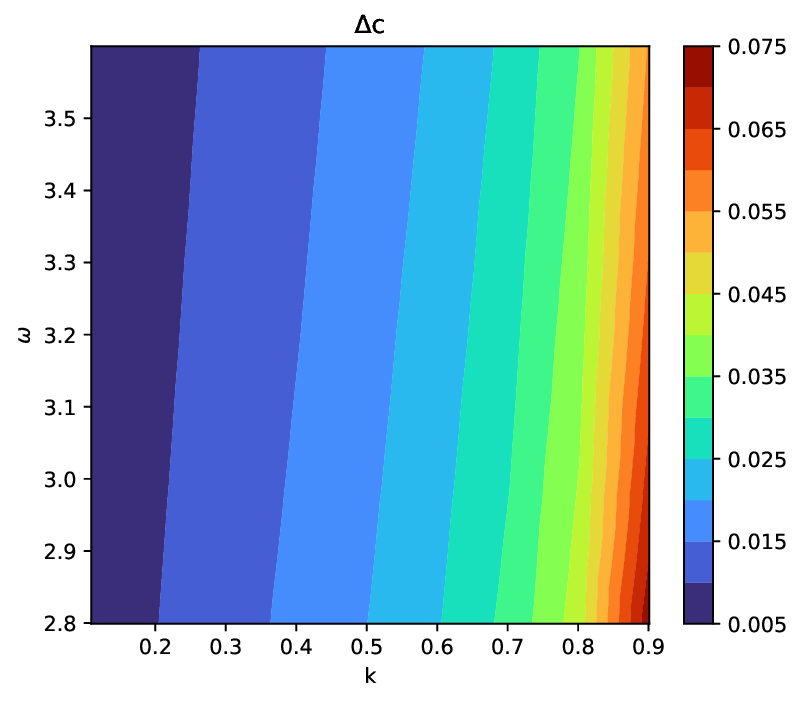}
	\end{minipage}
     \hfill
	\begin{minipage}{0.35\textwidth}
	\includegraphics[scale =0.6]{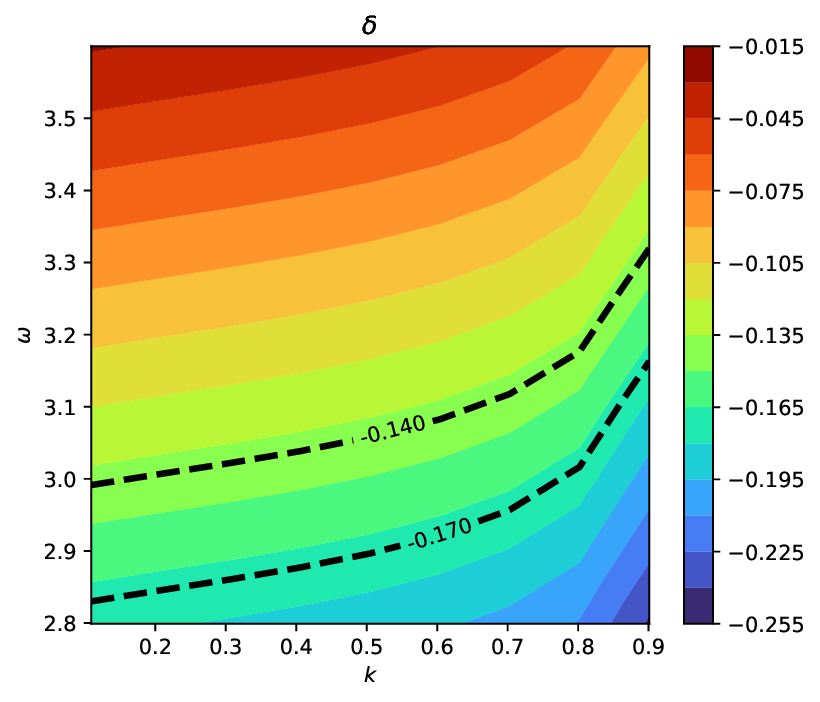}
	\end{minipage}
\caption{Contours of the deviation from circularity, $\Delta C$ (left panel), and the fractional shadow diameter deviation, $\delta$ (right panel), in the $(\omega,k)$ parameter space for fixed values of $a=0.9$ and $g=0.9$. The results are shown for an observer inclination angle of $17^\circ$.}
\label{Fig4}
\end{figure*}

\begin{figure*}
     \centering
     \begin{minipage}{0.4\textwidth}
	\includegraphics[scale =0.6]{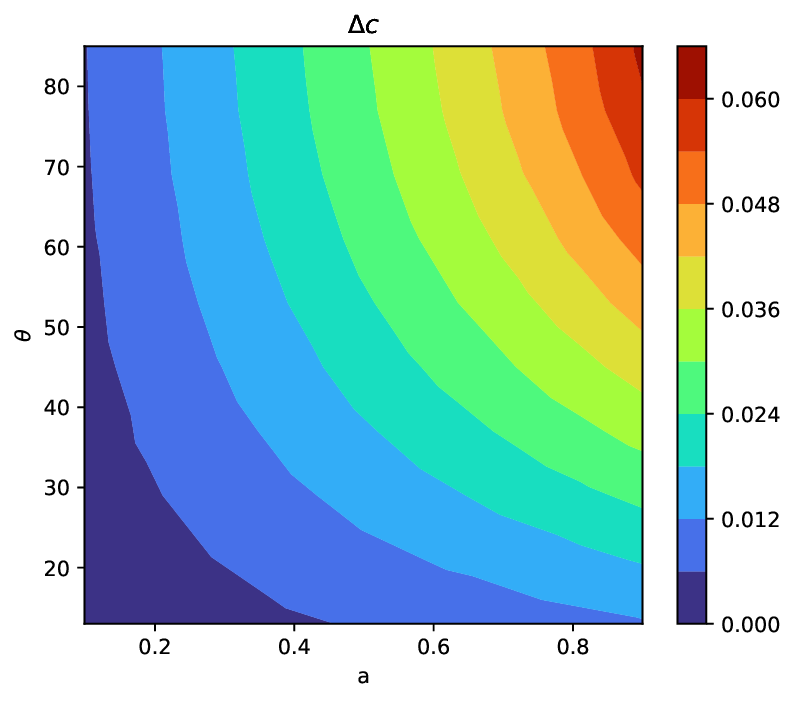}
	\end{minipage}
     \hfill
	\begin{minipage}{0.35\textwidth}
	\includegraphics[scale =0.6]{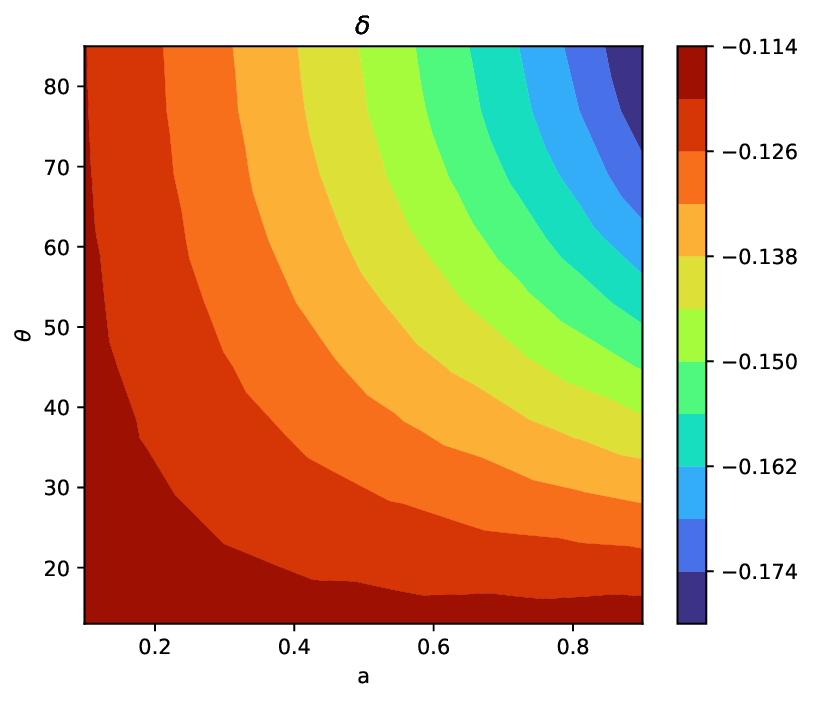}
	\end{minipage}
\caption{Contour plots of the deviation from circularity, $\Delta C$ (left panel), and the fractional shadow diameter deviation, $\delta$ (right panel), in the $(a,\theta)$ parameter space for fixed values of $k=0.4$, $g=0.25$, and $\omega=3.0$.}
\label{Fig5}
\end{figure*}

\begin{figure*}
     \centering
     \begin{minipage}{0.3\textwidth}
	\includegraphics[scale =0.6]{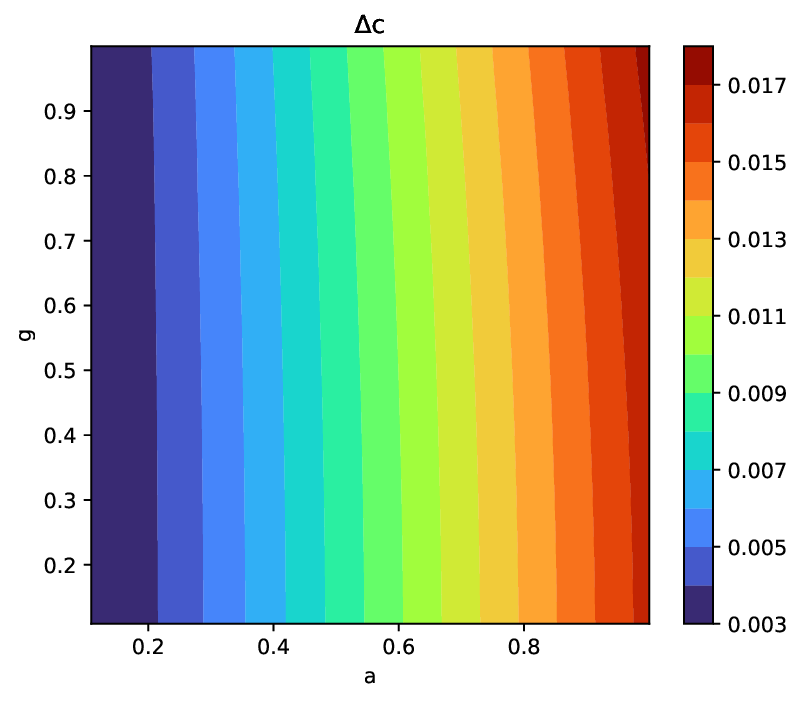}
	\end{minipage}
     \hfill
	\begin{minipage}{0.35\textwidth}
	\includegraphics[scale =0.6]{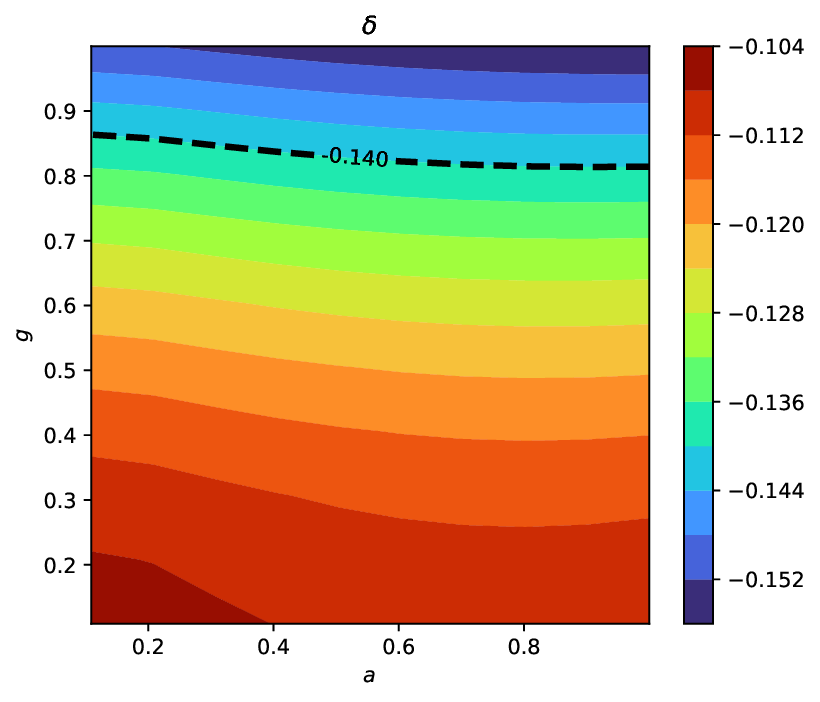}
	\end{minipage}
\caption{Contours of the deviation from circularity, $\Delta C$ (left panel), and the fractional shadow diameter deviation, $\delta$ (right panel), in the $(a,g)$ parameter space for fixed values of $\omega=3.0$ and $k=0.4$. The results are shown for an observer inclination angle of $17^\circ$.}
\label{Fig6}
\end{figure*}

To reduce the complexity of the analysis,the plasma parameter $k$ and the PFDM parameter $\omega$ were first constrained as they are the external environments that influence the shadow observables . We analysis  the shadow observables for the homogeneous plasma model. This choice was motivated by the fact that the radially inhomogeneous plasma was investigated under the near-equatorial approximation, making a comprehensive inclination-dependent parameter study physically inconsistent, whereas the general plasma model with both radial and angular dependence requires significantly more involved numerical computations that hinder a systematic exploration of the full parameter space.

The constrained values obtained from the homogeneous plasma model were therefore adopted for the subsequent analysis of the intrinsic black hole parameters.Figure 8 reveals the contour plots in the $k -\omega$ parameter space, We find that the deviation from circularity, $\Delta C$, increases monotonically with the plasma parameter, whereas the dependence on the PFDM parameter remains comparatively weaker. In contrast, the fractional diameter deviation, $\delta$, exhibits a combined dependence on both $k$ and $\omega$, with the EHT observational bounds ($\delta=-0.140$ and $\delta=-0.170$) restricting the physically admissible region of the parameter space we constraint the $\omega$ and k value above the highlighted contour for the subsequent analysis.

Using the constrained values of the plasma and PFDM parameters, the influence of the intrinsic black hole parameters, namely the spin parameter $a$, magnetic charge $g$, and observer inclination angle $\theta$, was subsequently investigated in Figure 9.

The resulting contour maps demonstrate that the deviation from circularity is predominantly governed by the spin parameter, while the magnetic charge produces only a comparatively weak modification. The observer inclination further amplifies the shadow asymmetry, leading to the largest deviations for rapidly rotating black holes viewed close to the equatorial plane. On the other hand, For Figure 10,we note that the fractional diameter deviation is found to be considerably more sensitive to the magnetic charge, thereby providing a stronger observational constraint on $g$. We note that both Deviation from circularity and fractional deviation parameter are both under permissible range of EHT observational data.

\section{Conclusion}

We consider a rotating regular Bardeen black hole embedded in a perfect fluid dark matter (PFDM) environment and immersed in a non-magnetized plasma medium. It is also assumed that the Plasma does not interact with the PFDM. Initial horizon analysis of Rotating Bardeen black hole with PFDM shows that the outer horizons is  distinguished to two regions namely, lower bound PFDM range and upper bound PFDM range.The shadows are studied in both ranges.

We consider two forms of plasma distributions for our shadow analysis. In the first case,the plasma frequency is assumed to be purely radial, with two models: a homogeneous profile \(n = \sqrt{1 - k}\), and an inhomogeneous one given by \(n(r) = \sqrt{1 - \frac{k}{r}}\). In the second case,  the plasma frequency depends on both radial and angular coordinates, defined through specific functional forms: \(f_r(r) = \psi_c^2 \sqrt{M^3 r}\) with \(f_\theta(\theta) = 0\), and \(f_r(r) = 0\) with \(f_\theta(\theta) = \psi_c^2 M^2 (1 + 2 \sin^2\theta)\).

The shadow characteristics are found to depend sensitively on the spin parameter $a$, magnetic monopole charge $g$, PFDM parameter $\omega$, and plasma strength parameter $k$. The rotation of the black hole enhances the asymmetry of the shadow through frame dragging, while increasing the magnetic charge further modifies the spacetime geometry, leading to noticeable changes in both the size and deformation of the shadow.The PFDM parameter introduces a non-linear effect: the shadow size decreases with increasing \(\omega\) when \(\omega < \omega_c\), but beyond the critical point \(\omega > \omega_c\), the shadow begins to expand again, consistent with the influence of effective mass discussed earlier.

The presence of plasma alters photon trajectories by changing the refractive index of the medium. Our analysis further reveals that this effect varies depending on the nature of the plasma distribution. In the Homogeneous plasma distribution,the black hole distortion is significant with presence of plasma, and the effect decreases as the plasma density reduces with distance in the inhomogeneous distribution

The general plasma distributions exhibit distinct effects on the black hole shadow. While the radial plasma profile has only a weak influence on the shadow size and shape, with spin-induced distortion being more pronounced in the lower-bound PFDM regime, the angular plasma profile primarily alters the shadow size rather than its shape. Increasing the plasma frequency reduces the shadow radius, an effect that becomes significantly stronger in the upper-bound PFDM regime. Thus, the radial and angular plasma distributions predominantly affect the shadow distortion and shadow size, respectively.

We also analyze the effective potential governing photon motion in the vicinity of the black hole. As the plasma strength parameter increases, the peak of the effective potential shifts toward smaller radial distances while its magnitude increases. This behavior indicates that the presence of plasma modifies the effective refractive environment through which photons propagate, thereby changing the location of the unstable circular photon orbits. These variations are reflected in the corresponding changes in the black hole shadow.

Further we  constraint the $\omega - k $Parameter space with EHT observational data on the deviation from circularity and fractional diameter deviation, thereby identifying the physically admissible region of the parameter space. Using the constraints, our analysis shows that the deviation from circularity is primarily governed by the black hole spin and observer inclination, whereas the magnetic charge has a comparatively weaker influence on the shadow shape. In contrast, the fractional diameter deviation is found to be more sensitive to the magnetic charge, making it an effective observable for constraining the nonlinear electrodynamic parameter. Overall, the results demonstrate that combining multiple shadow observables with current EHT constraints provides a robust framework for simultaneously constraining both the environmental parameters and the intrinsic properties of regular black holes, thereby offering a more reliable interpretation of future high-resolution black hole observations.

\end{document}